\documentclass[1p]{elsarticle}

\usepackage{amssymb,amsmath}
\usepackage{graphicx}
\usepackage{pgf,pgfarrows,pgfnodes,pgfautomata,pgfheaps}
\usepackage{multicol}
\newcommand{\etal}{\MakeLowercase{\textit{et al. }}}

\begin{document}
\author[ucm,uc]{M.~Monasor\corref{cor1}} \ead{monasor@uchicago.edu}
\author[ucm]{J.~R.~V\'azquez}
\author[ucm]{D.~Garc\'ia-Pinto} 
\author[ucm]{F.~Arqueros}

\cortext[cor1]{Corresponding author} 

\address[ucm]{Departamento de F\'isica At\'omica, Molecular y Nuclear, Universidad Complutense de Madrid, Madrid, Spain} 
\address[uc]{Department of Astronomy \& Astrophysics, University of Chicago \& Kavli Institute for Cosmological Physics, Chicago, U.S.A.}

\title{\bf The impact of the fluorescence yield on the reconstructed shower parameters of ultra-high energy cosmic rays}

\begin{abstract}
An accurate knowledge of the fluorescence yield and its dependence on atmospheric properties such as pressure, temperature or humidity is essential to obtain a reliable measurement of the primary energy of cosmic rays in experiments using the fluorescence technique. In this work, several sets of fluorescence yield data (i.e. absolute value and quenching parameters) are described and compared. A simple procedure to study the effect of the assumed fluorescence yield on the reconstructed shower parameters (energy and shower maximum depth) as a function of the primary features has been developed. As an application, the effect of water vapor and temperature dependence of the collisional cross section on the fluorescence yield and its impact on the reconstruction of primary energy and shower maximum depth has been studied.
\end{abstract}

\maketitle

\section{Introduction}
Fluorescence telescopes record the longitudinal profile of extensive air showers induced by very energetic cosmic rays through the
detection of the fluorescence light generated by secondary charged particles. This technique allows an accurate determination of
the shower maximum depth $X_{\rm{max}}$. In addition, since the fluorescence intensity is proportional to the deposited energy,
the integration in depth of the fluorescence profile allows a calorimetric determination of the shower energy. The total primary energy 
is later obtained by applying a correction accounting for the so-called missing energy. A key parameter for the reconstruction of the calorimetric energy is the fluorescence yield (FY), defined as the number of fluorescence photons emitted per unit deposited energy. The FY which depends on the atmospheric parameters (i.e. pressure $P$, temperature $T$, humidity) is measured in dedicated laboratory experiments. Presently one of the largest
contributions to the total systematic error in the primary energy comes from the absolute value of this parameter. Also the
uncertainties in the dependence of the FY on atmospheric parameters might have a non-negligible effect on both primary energy
and $X_{\rm{max}}$ measurements.
\par
In this paper a simple procedure to study the effect of the assumed fluorescence yield on the reconstructed shower parameters
will be shown. Preliminary results were published in~\cite{ICRC_09_FY}. 
In section~\ref{sec:AFY} a brief summary of the FY
properties will be presented. In the next section the features of some data sets of FY commonly used will be described and
compared. A simple method to evaluate the effect of the FY selection on shower reconstruction is shown in section 4. 
The shower development will be described by a Gaisser-Hillas profile. The effect of a variation
in the FY (including its atmospheric dependence) is a change in the reconstructed longitudinal development of the deposited
energy which can be easily evaluated. Results for the dependence of both reconstructed $X_{\rm{max}}$ and primary energy on the
FY selection will be presented in section~\ref{sec:results}. 
\par
Keilhauer and Unger~\cite{ICRC_09_Bianca} have carried out a similar study focused on estimating the influence of the atmospheric conditions on the reconstructed shower parameters. However our approach is somewhat different since we only need the Gaisser-Hillas profile, while in~\cite{ICRC_09_Bianca} simulated showers are used for the analysis.

\section{Air-Fluorescence Yield}
\label{sec:AFY} Air-fluorescence in the near UV range is basically produced by the de-excitation of atmospheric nitrogen
molecules excited by the shower electrons. The spectrum of fluorescence consists of a set of molecular bands represented by
their wavelengths $\lambda$ and therefore the total fluorescence yield $Y$ in a given wavelength interval $\Delta\lambda$ can be
obtained by adding up the contributions of all the molecular bands $Y_{\lambda}$.
\begin{equation}
Y(X) = \sum_{\Delta\lambda} Y_\lambda(X) \label{eq:total_FY}\,.
\end{equation}
Excited molecules can also decay by collisions with an environmental molecule. Because of this effect, the FY in the absence of
quenching $Y_\lambda^0$ (that is, at null pressure) is reduced by the so-called Stern-Volmer factor.
\begin{equation}
\label{eq:FY} Y_{\lambda}(P,T) = \frac{Y_{\lambda}^0}{1+P/P'({\lambda, T})}\,.
\end{equation}
The dependence of $Y_{\lambda}$ on atmospheric conditions can be described by a single parameter, the characteristic
pressure $P'$ which is defined as the pressure for which both radiative and collisional de-excitation have the same probability.
In general $P'$ contains a contribution of all possible quenchers $i$ (i.e., N$_2$, O$_2$, H$_2$O).
\begin{equation}
\label{eq:1_P} \frac{1}{P'} = \sum_i{\frac{f_i}{P'_i}}\,,\quad P'_i = \frac{kT}{\tau}\frac{1}{\sigma_{{\rm N}i} \bar{v}_{{\rm N}i}}\,,\quad
\bar{v}_{{\rm N}i}=\sqrt{\frac{8kT}{\pi \mu_{{\rm N}i}}}\,.
\end{equation}
In the above expressions $f_i$ is the fraction of molecules of type $i$ in the mixture, $\sigma_{{\rm N}i}$ is the collisional cross
section which depends on the particular band, and $v_{Ni}$ and $\mu_{Ni}$ are the relative velocity and reduced mass of the two
body system N-i respectively; $k$ is the Boltzman constant and $\tau$ the radiative lifetime of the corresponding level.
\par
A number of experimental results of the FY in dry air are available (a review is presented in~\cite{5th_FW_SP} and a critical comparison of the absolute measurements can be found in \cite{rosado}). Usually the absolute values of $Y_{\lambda}$ are measured at atmospheric pressure and room temperature and the $P'$ parameters are evaluated
at the same temperature. Equation~(\ref{eq:FY}) allows the determination of $Y_{\lambda}^0$ and the dependence of the FY with
pressure. Unfortunately measurements of the $P'$ parameters show large discrepancies and therefore the $Y_{\lambda}^0$ parameter
is not suitable for comparison between different experiments.
\par
According to expression~(\ref{eq:1_P}), the $P'$ value for dry air can be expressed as
\begin{equation}
\frac{1}{P'_{\rm air}(\lambda,T)}=\frac{f_{\rm O}}{P'_{\rm O}(\lambda,T)} + \frac{f_{\rm N}}{P'_{\rm N}(\lambda,T)}\,,
\label{eq:dry_air}
\end{equation}
\noindent where $f_{\rm O}=0.21$ and $f_{\rm N}=0.79$ are the fraction of oxygen and nitrogen molecules in air respectively. Measurements of $P'_{\rm O}(\lambda)$ and $P'_{\rm N}(\lambda)$ are also
available~\cite{Tilo}.
\subsection{Humidity dependence}
\label{sec:humidity} Water molecules are also effective quenchers for excited $N_2$ and therefore fluorescence emission is
partly suppressed in humid air. The characteristic pressure for humid air $P'_{\rm hum}(\lambda,T)$ can be calculated from
equation~(\ref{eq:1_P}) by including the water contribution. The relationship between the characteristic pressures of dry
$P'_{\rm dry}(\lambda, T)$ and humid air $P'_{\rm hum}(\lambda, T)$ is
\begin{equation}
\frac{1}{P'_{\rm hum}(\lambda,T)}=\frac{1}{P'_{\rm dry}(\lambda, T)} \left(1- \frac{P_{\rm w}}{P}\right) + \frac{P_{\rm
w}}{P}\frac{1}{P'_{\rm w}(\lambda, T)}\,,
\end{equation}
\noindent where $P_{\rm w}$ is the partial pressure of water and $P'_{\rm w}$ is the characteristic pressure for fluorescence
quenching by collisions with water molecules.
\par
Measurements of $P'_{\rm w}$ values have been performed by several authors. The AIRFLY collaboration has published results for
four bands at 293 K~\cite{temp_cross_sec}. In ref.~\cite{Tilo} experimental values of $P'_{\rm w}$ for three relevant bands are
also reported. More details can be found in~\cite{5th_FW_SP}.
\subsection{Temperature dependence}
\label{sec:temperature}
The $T$ dependence of the fluorescence yield is given by equations~(\ref{eq:FY}) and~(\ref{eq:1_P}).
Neglecting the dependence of the collisional cross section on the kinetic energy of the colliders, $P'$ is proportional to
$\sqrt{T}$. Assuming a measurement of the characteristic pressure at a reference temperature $T_0$, the $T$ dependence of
$P'_i(\lambda,T)$ is given by
\begin{equation}
P'_i(\lambda,T) = P'_i(\lambda,T_0)\sqrt{\frac{T}{T_0}} \label{eq:pprime_varT}\,.
\end{equation}
However, as it is well known, the average collisional cross section follows a behavior which can be described by a power law ($\sim
T^{\alpha}$)~\cite{cross_section}. Therefore equation~(\ref{eq:pprime_varT}) becomes
\begin{equation}
P'_i(\lambda,T) = P'_i(\lambda,T_0)\sqrt{\frac{T}{T_0}} \cdot \frac{T_0^{\alpha^i_\lambda}}{T^{\alpha^i_\lambda}} =
P'_i(\lambda,T_0)\left(\frac{T_0}{T}\right)^{\alpha^i_\lambda-1/2} \label{eq:pprime_mod}\,.
\end{equation}
The dependence of the FY with both $T$ and $P$ can be easily predicted by combining equations~(\ref{eq:FY}), (\ref{eq:1_P}) and (\ref{eq:pprime_mod}). The temperature dependence of the collisional cross section for dry air has been measured recently by the AIRFLY collaboration and the corresponding values of $\alpha_\lambda$ parameter for four molecular bands have been published~\cite{temp_cross_sec}.  In principle, this parameter depends on the quencher (nitrogen, oxygen, water). However independent measurements of $\alpha^i_\lambda$ for each air component is difficult and usually it is assumed that $\alpha^i_\lambda$ takes the same value for all components. 
On the other hand $\alpha_\lambda$ values for water are not presently available. However, as we will show below all these uncertainties and approximations have a marginal impact on the shower reconstruction.
\section{Air-fluorescence yield data}
\label{sec:AFY_data}
\subsection{Required information}
\label{sec:info} The reconstruction of the shower parameters requires the following information on the fluorescence yield.
\begin{itemize}
\item[1)] The absolute value in dry air at a given pressure and temperature for all bands within the spectral range of the
telescope. As an alternative, the absolute value for a reference transition (e.g. 337 nm) and the relative intensities of the
bands at a given pressure will also provide the total fluorescence yield.
\item[2)] The values of $P'$ in dry air at a reference temperature for all bands. Measurements of $P'_{\rm N}(\lambda,T_0)$ and
$P'_{\rm O}(\lambda,T_0)$ could be also used for this purpose.
\item[3)] The values of $P'_{\rm w}(\lambda, T_0)$ for all wavelengths.
\item[4)] The $T$ dependence of $\sigma_{{\rm N}i}$ for all wavelengths. In principle $\alpha_{\lambda}$ values for each quencher
would be needed, however this information is, at present, rather limited.
\end{itemize}
A data set containing 1) and 2) allows us to determine the total FY for dry air as a function of $P$ at the reference temperature
$T_0$. Neglecting the $T$ dependence of $\sigma_{Ni}$, the total FY can be calculated at any $P$ and $T$ condition. Adding 3),
the air-fluorescence yield can be evaluated for humid air. Finally, 4) provides a more accurate extrapolation at temperatures
far from the reference one (i.e. at high altitude).
\par
Apart from pioneering works~\cite{NIM_pioneering}, in the last 10 years several measurements of the FY have been carried out in
laboratory experiments injecting accelerated electrons into air targets~\cite{5th_FW_SP}. Nowadays three data sets combining
some of these measurements are mainly being used in cosmic ray experiments using fluorescence telescopes. They are those of Kakimoto-Bunner, Nagano and the combination Nagano-AIRFLY. These data sets are described below. In all
of them the humidity effect is neglected and the $T$ dependence of quenching is calculated assuming a constant collisional cross
section. The effect of these approximations on the fluorescence yield is not negligible and will be also evaluated in this
section.
\subsection{Kakimoto-Bunner (K-B)}
\label{sec:K-B}
This FY data set which was used by the HiRes collaboration in 2001~\cite{dawson} is a combination of fluorescence yield measurements of Kakimoto et al.~\cite{kakimoto} and the relative
intensities of the molecular bands reported by Bunner~\cite{bunner_thesis}.
\par
Kakimoto et al. measured the absolute value of the air-fluorescence yield expressed
in photons per meter $\epsilon_{\lambda}$ for the three primary molecular bands of nitrogen (337, 357
and 391 nm) at 1000 hPa and 288 K. The remaining fluorescence bands are found to contribute with about a 30\% of the total
fluorescence photons when the filter employed in the HiRes telescopes (bandpass filter between 295 and 405 nm) is considered. As pointed out by~\cite{dawson}, these data translate into an absolute FY contribution of 0.46 ph/m for these bands. The
remaining fluorescence spectrum (13 bands) is distributed according to the relative intensities collected by Bunner~\cite{bunner_thesis}.
\par
Kakimoto et al. calculated the fluorescence efficiency
of the primary bands (defined as the fraction of deposited energy emitted as
fluorescence radiation) $\phi_\lambda$ as
\begin{equation}
\phi_\lambda = \frac{\epsilon_{\lambda}}{\rho\left({\rm d}E/{\rm d}X\right)_{\rm dep}}\cdot\frac{hc}{\lambda}\,,
\end{equation}
\noindent where $h$ is the Planck constant, $c$ is the speed of light, ${\left({\rm d}E/{\rm d}X\right)_{\rm dep}}$ is the energy deposited
by the electron per unit of atmospheric depth, and $\rho$ is the air density.
\par
Experimental results of pressure dependence for the primary bands are also reported by Kakimoto et al. In that work the FY is
parameterized as
\begin{equation}
\epsilon_\lambda (P,T)= \frac{\rho A_\lambda}{1+\rho B_\lambda \sqrt{T}}\,.
\end{equation}
In this expression $A_\lambda$ is the number of photons emitted per electron and unit column density (g/cm$^2$) in the absence
of quenching (i.e. $P$=0) and the $B_\lambda$ parameter is related with the characteristic pressure through~(\ref{eq:FY}) and
(\ref{eq:1_P}) by
\begin{equation}
 B_\lambda = \frac{R_{\rm gas}\sqrt{T}}{P'_{\rm air}(\lambda,T)}\,,
\end{equation}
\noindent where $R_{\rm gas}$ is the specific gas constant.
\par
The implementation of $\phi_\lambda$ and $B_\lambda$ in equation~(\ref{eq:1_P}) can be easily performed. Firstly, the
fluorescence yield is related with the fluorescence efficiency by
\begin{equation}
Y_\lambda = \phi_\lambda\cdot\frac{\lambda}{hc} \label{phi}\,.
\end{equation}
Therefore the $Y_\lambda^0$ values can
be obtained from the fluorescence efficiency at $P_0$=1000 hPa and the $B_\lambda$ parameters at $T_0$ = 288 K from
\begin{equation}
Y_\lambda^0 = \phi_\lambda(P_0)\frac{\lambda}{{h}c}\left(1+ \frac{P_0 B_\lambda}{R_{\rm gas} \sqrt{T_0}}\right)\,.
\end{equation}
Finally, the values of $P'_\lambda$ inferred from the reported $B_\lambda$ parameters of Kakimoto \etal for the 337 nm band are used for
all 2P components while that of the 391 nm is used for the 1N system.
\subsection{Nagano}
\label{sec:Nagano}
Nagano {\it et al.}~\cite{nagano} provided experimental results of $\epsilon_{\lambda}$ in dry air for 15
nitrogen bands at 1013 hPa and 293 K as well as the corresponding $P'_\lambda$ values at the same temperature. Using these
measurements, the authors calculate the fluorescence efficiency at null pressure $\phi^0_\lambda$. The corresponding $Y_\lambda^0$ values can be easily obtained
applying expression~(\ref{phi}). Notice that both Kakimoto \etal  and Nagano \etal  assume for this calculation that the energy lost by the electron is fully deposited in the field of view of the experimental set-up. Implication of this approximation has been discussed in \cite{rosado},~\cite{moriond} and~\cite{njp_nagano}. 
\subsection{Nagano-AIRFLY (N-A)}
In this data set the absolute scale is given by the $\phi_{337}^0$ value reported by Nagano \etal for the 337 nm band which translates into a $Y_{337}$ value of 6.38 ph/MeV at $P_0$=800 hPa and $T_0$=293 K. Relative intensities $ I_\lambda$ at $T_0$ and $P_0$ for 34 wavelengths and the corresponding $P'(\lambda)$ values in dry air at 293K reported by AIRFLY~\cite{AIRFLY_2007}  are used to get a complete set. The corresponding $Y_\lambda^0$ for any wavelength is thus obtained as
\begin{equation}
Y_\lambda^0 = Y_{337}(P_0, T_0) \cdot I_\lambda(P_0, T_0) \cdot \left(1+\frac{P_0}{P'_{\rm air}(\lambda,T_0)}\right)\,.
\end{equation}
This data set is presently used by the Pierre Auger Observatory~\cite{auger_spectrum}.
\subsection{Comparison of data sets}
\label{sec:comparison} 
\begin{figure}
\centering
\includegraphics[height=0.5\textwidth]{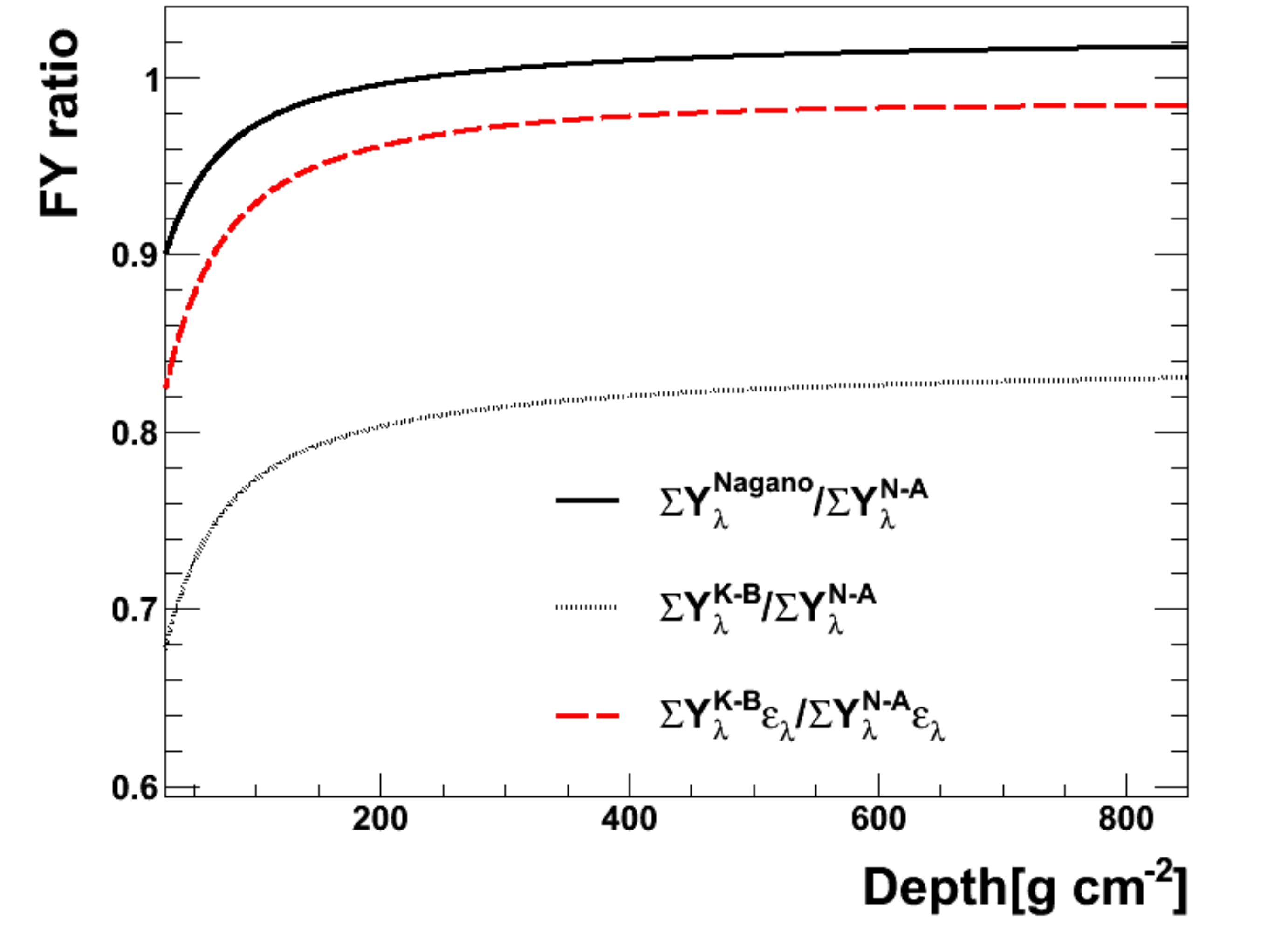}
\caption{\footnotesize{Ratio of Nagano and Kakimoto-Bunner fluorescence yields to that of Nagano-AIRFLY versus atmospheric depth. The ratio K-B/N-A of $\Sigma Y_\lambda(X) \varepsilon_\lambda$, also displayed, indicates that the effect of the 
optical efficiency (filter and telescope) reduces significantly the large disagreement in the total FY of both data sets.
}}
\label{fig:fy2}
\end{figure}

The FY for the above described data sets has been compared. 
In figure~\ref{fig:fy2} the ratios $Y_{\rm Nagano}/Y_{\rm N-A}$ and $Y_{\rm K-B}/Y_{\rm N-A}$ 
against atmospheric depth are shown. The large deviations observed at low depth are due to discrepancies in
the $P'$ values in the data sets. Obviously these discrepancies have no relevant impact in shower reconstruction since in our energy range they take place before the development of the shower. On the other hand, at larger depths differences around 20\% between N-A and K-B data sets are found while a relative small discrepancy of around 2\% is observed between N-A and Nagano descriptions.
\par
In practice the comparison between FY data sets should include other ingredients. Fluorescence photons generated by cosmic ray showers are detected with ground telescopes located at a large distance from the emission point and thus suffering a non-negligible attenuation which is wavelength dependent. In addition, the optical systems of the telescope, including the filter, the mirror and the PMT photocathode have a response which is also wavelength dependent. Therefore the relative differences between two FY data sets are not directly translated to the energy reconstruction since the relative intensities of the molecular bands are modified by the detector efficiency. For instance, discrepancies between two data sets which only affect wavelength regions for which the filter has a low transmittance  might have a marginal impact on energy reconstruction although the discrepancy in total FY is large. The parameter more closely related with possible deviations in the shower reconstruction is the product $Y_\lambda(X) \varepsilon_\lambda T_\lambda(X,X_0)$ where $\varepsilon_\lambda$ is the optical efficiency of the telescopes including all components and $T_\lambda(X,X_0)$ represents the atmospheric transmittance between the emission point and the telescope location.
\par
\begin{figure}[htb]
\includegraphics[width=0.49\textwidth]{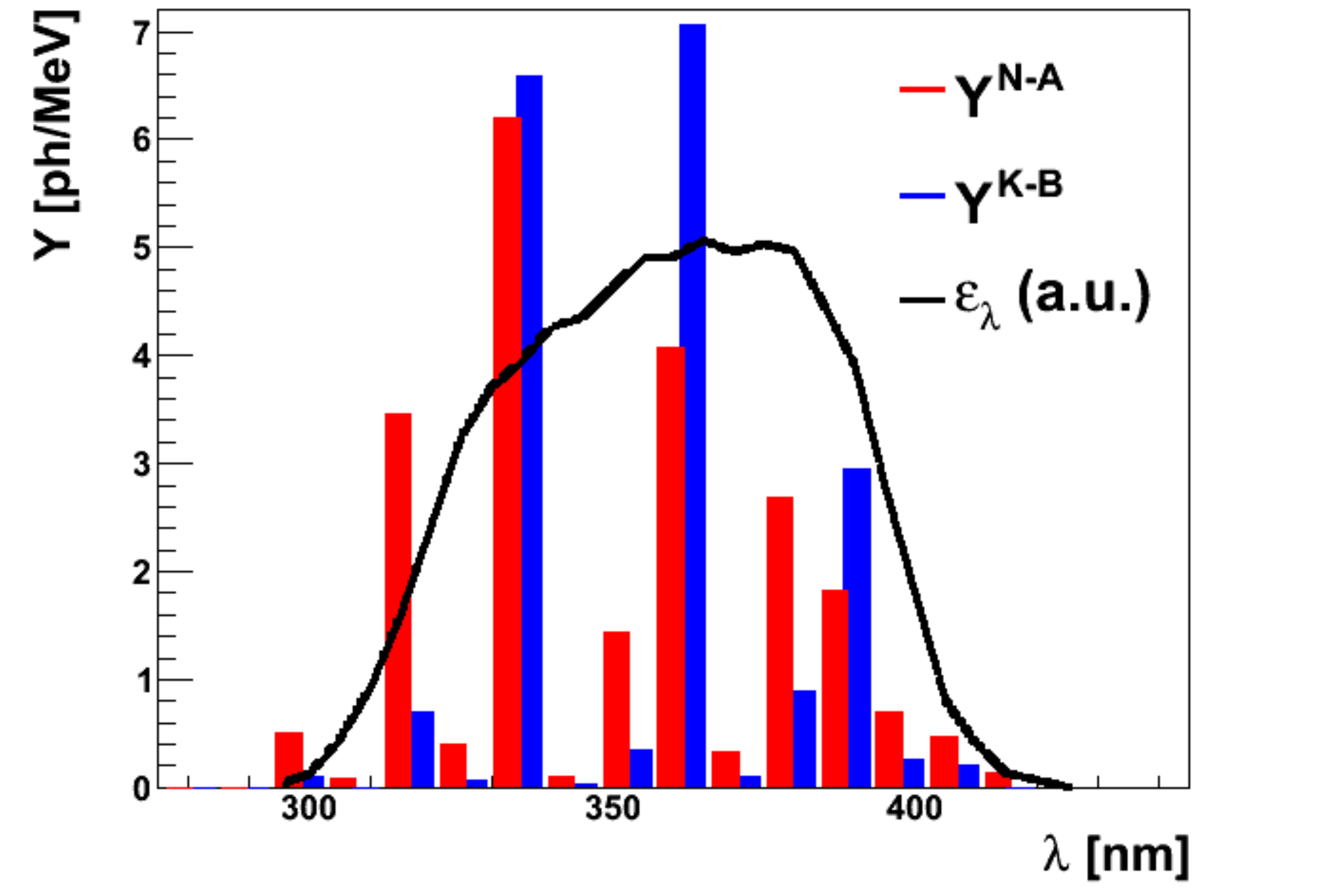}
\includegraphics[width=0.49\textwidth]{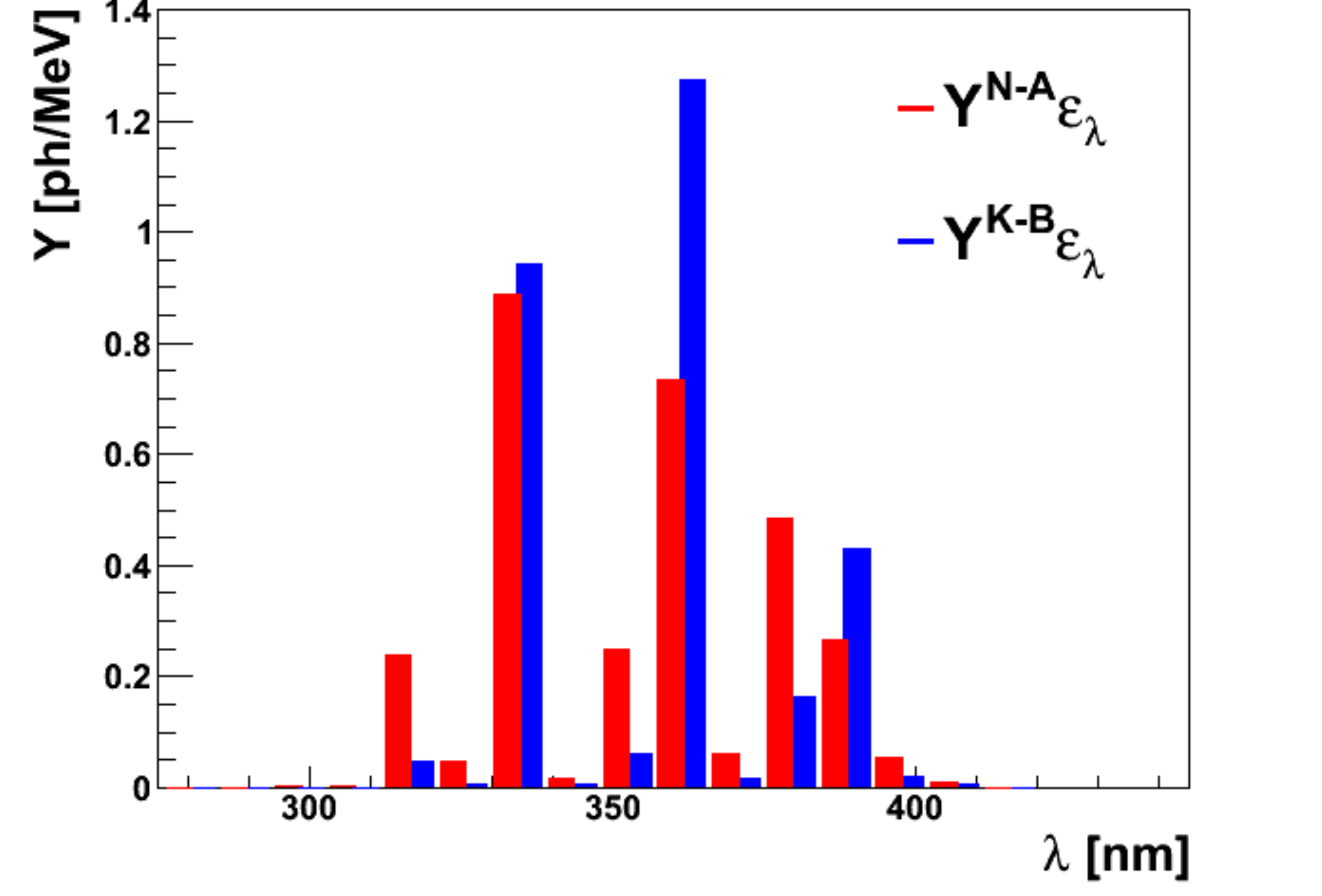}
\caption{\footnotesize{Fluorescence yield in absolute units versus wavelength for the N-A and K-B data sets (left). Since K-B intensities are given in 9 nm bins~\cite{dawson}, the N-A fluorescence yields have been recalculated for the same wavelength bins for a better comparison. The effect of the Auger optical efficiency (black line) on the FY of both data sets is illustrated (right).}}\label{fig:fy3}
\end{figure}

The effect of the optical filter is particularly relevant in the comparison between K-B and N-A data sets. As shown in 
figure~\ref{fig:fy2} the large disagreement between them nearly disappears when the efficiency of the optical system of the Auger telescopes \cite{auger_filter} is included. Figure~\ref{fig:fy3} shows the fluorescence yield for both data sets (left) and the effect of the optical efficiency (right). Note that the large discrepancy in the bin centered at 317 nm is smoothed by the strong absorption of the filter at this wavelength. However the discrepancies at wavelengths with good transmission, and thus less sensitive to the filter, nearly compensate each other. 
\par
In a similar way, it has been shown~\cite{nagano} that the large disagreement between the Nagano and K-B data sets are significantly reduced when  the optical efficiency of HiRes is included. 
\subsection{Effect of temperature and humidity on the FY}
\label{sec:Effects_T_h}
\begin{figure}
\includegraphics[width=1.0\textwidth]{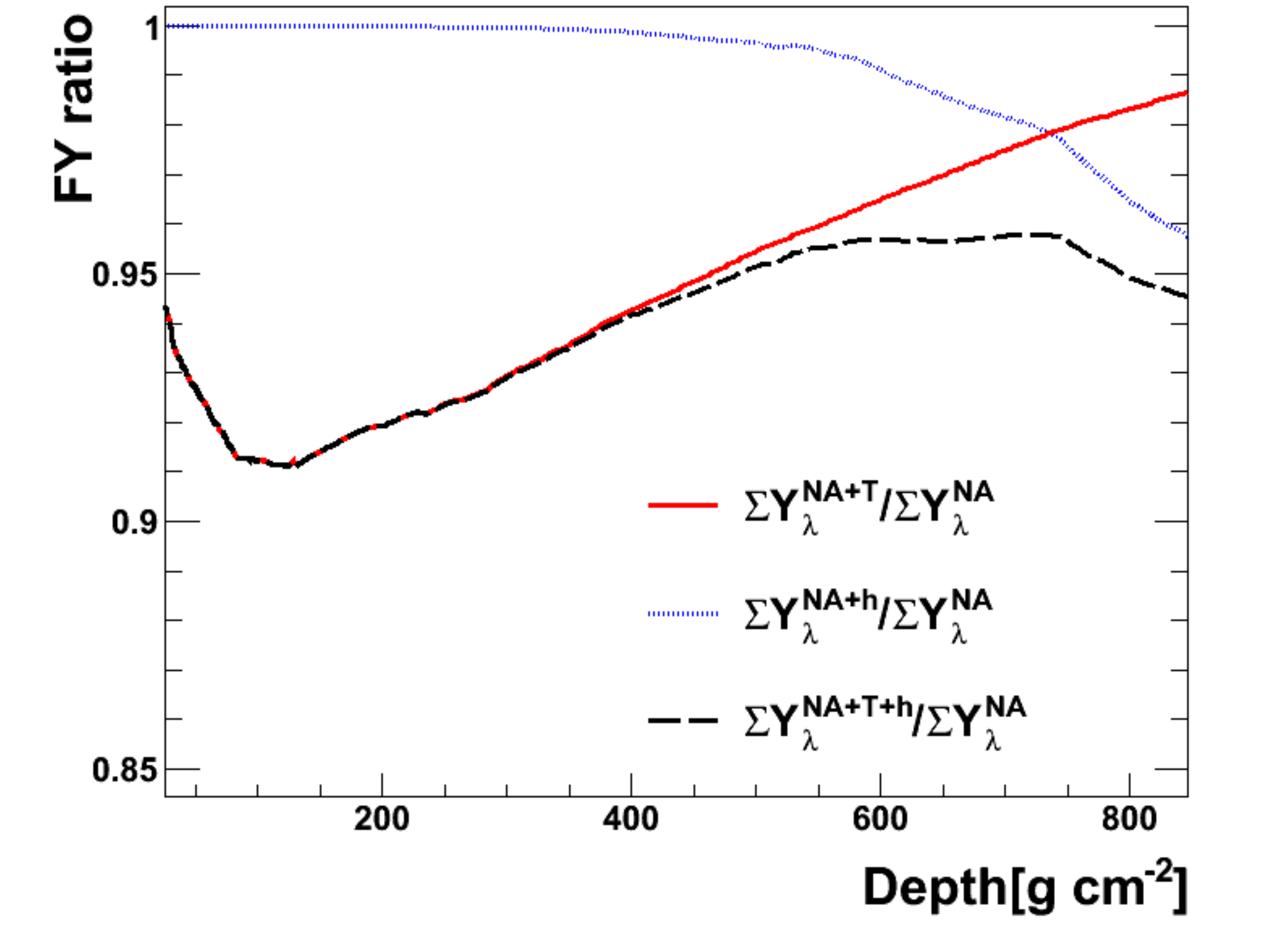}
\caption{\footnotesize{Effect of the humidity and a T-dependent collisional cross section on the N-A data set. In this plot the ratio of fluorescence yields for the month of December versus atmospheric depth is shown.}}\label{fig:fy_hT}
\end{figure}

The above described data sets do not include the contribution of quenching by water molecules in humid air.
In addition, the temperature dependence of the characteristic pressure is assumed to follow a $\sqrt{T}$ law (i.e.
$\alpha_\lambda$ is assumed to be zero). The effect of these approximations on the FY profile has been evaluated using the 
monthly average description for the atmosphere measured at the Auger site~\cite{keilhauer_atm}. As an example, in figure \ref{fig:fy_hT} 
the effect of including the humidity and temperature corrections (December month) in the N-A data set is shown against atmospheric depth. 
For these calculations the quenching parameters of~\cite{temp_cross_sec} have been used. 
As expected, humidity decreases the fluorescence yield at large depth (i.e. near ground) while the effect of the $T$ term is only relevant at high altitudes. Neglecting the possible dependence of the collisional cross section N$_2^*$-H$_2$O with $T$ (i.e. assuming $\alpha_\lambda$ is zero for water), the total effect of humidity and temperature can be easily computed. The result is also shown in figure~\ref{fig:fy_hT}.
A similar behavior has been found in a study carried out in \cite{bianca_NIM_A} using several sets of fluorescence yield data including the temperature and humidity dependence of quenching from various authors. 
\par
As expected, neither the optical efficiency of the detector nor the atmospheric transmission introduce any
significant effect in these results since atmospheric conditions (i.e. temperature and humidity) has a negligible impact 
on the fluorescence spectrum.
\section{Method}
\label{sec:AFY_method}
In this note we propose a simple procedure to evaluate the effect of the fluorescence yield selection on the
reconstruction of shower parameters. Assuming a certain longitudinal profile of deposited energy ${\rm d}E(X)/{\rm d}X$, the number of
fluorescence photons generated per unit atmospheric depth is determined by the fluorescence yield in the wavelength interval of
the telescope.
\begin{equation}
\frac{{\rm d} n^{\rm gen}_\gamma}{{\rm d}X} = \frac{{\rm d}E}{{\rm d}X} Y(X)\,.
\end{equation}
\par
The total calorimetric energy $E$ can be obtained from the integral of the longitudinal profile and therefore
\begin{equation}
E = \int_0^\infty \frac{1}{Y(X)}\frac{{\rm d} n^{\rm gen}_\gamma}{{\rm d} X}{\rm d} X\,. \label{Ecal}
\end{equation}
\par
If the FY assumption is now changed to a new data set $Y'(X)$, the profile of deposited energy consistent with the fluorescence
profile is
\begin{equation}
\frac{{\rm d}E'}{{\rm d}X} = \frac{{\rm d}E}{{\rm d}X} \, \frac{Y(X)}{Y'(X)}\,, \label{eq:new_dep}
\end{equation}
\noindent and thus the calorimetric energy associated to the new FY selection will be
\begin{equation}
E'= \int_0^\infty   \frac{{\rm d} E}{{\rm d}X} \, \frac{Y(X)}{Y'(X)}{\rm d} X \,.\label{Ecal_mod2}
\end{equation}
Therefore, in principle, the effect of changing the FY selection on the primary energy can be evaluated by comparing
(\ref{Ecal}) and (\ref{Ecal_mod2}). On the other hand, the effect on the $X_{\rm max}$ value can be obtained by comparing the
shape of the ${\rm d}E/{\rm d}X$ and ${\rm d}E'/{\rm d}X$ profiles.
\par
Only a small fraction of the fluorescence photons generated by the shower reaches the PMT camera of the fluorescence detector.
In the first place, the number of photons is strongly reduced by a factor $A/(4\pi R^2(X))$ where $A$ is the area of the telescope
and $R$ is the distance from the emission point to the telescope location. 
In addition, as previously pointed out,
the atmosphere scatters a non-negligible fraction of photons on their way to the telescope. Finally the optical elements of the telescope also
absorb a certain fraction of those photons reaching the telescope window. Both atmospheric transmission and optical efficiency are wavelength
dependent. Therefore the profile of deposited energy is calculated from the profile of observed photons using the expression
\begin{equation}
\frac{{\rm d}E}{{\rm d}X} = \frac{{\rm d}n^{\rm obs}_\gamma}{{\rm d}X}\cdot \frac{4\pi R(X)^2}{A} \cdot \frac{1}{\sum_{\Delta
\lambda} \varepsilon_\lambda \cdot T_\lambda(X) \cdot Y_\lambda(X)}\,,
\end{equation}
\noindent and thus, relationship (\ref{eq:new_dep}) becomes
\begin{equation}
\frac{{\rm d}E'}{{\rm d}X} = \frac{{\rm d}E}{{\rm d}X} \, \frac{\sum Y_\lambda(X)\cdot \varepsilon_\lambda \cdot T_\lambda(X)}{\sum
Y'_\lambda(X)\cdot \varepsilon_\lambda \cdot T_\lambda(X)} \,.\label{eq:new_dep_eff}
\end{equation}
Therefore the shower maximum depth has to be evaluated using (\ref{eq:new_dep_eff}) instead of (\ref{eq:new_dep}) and the
calorimetric energy $E'$ inferred from the new FY selection is given by
\begin{equation}
E' = \int_0^\infty   \frac{{\rm d} E}{{\rm d}X} \frac{\sum Y_\lambda(X)\cdot \varepsilon_\lambda \cdot T_\lambda(X)}{\sum
Y'_\lambda(X)\cdot \varepsilon_\lambda \cdot T_\lambda(X)}{\rm d} X \,. \label{Ecal_mod3}
\end{equation}
The variation of both $X_{\rm max}$ and
$E$ due to a change in the FY selection is not sensitive to the fine details
of the longitudinal development of ${\rm d}E/{\rm d}X$ and thus instead of using real data or simulated showers it is easier and more convenient for this study to describe the longitudinal development of deposited energy by a Gaisser-Hillas (GH) profile~\cite{gaisser-hillas} given by
\begin{equation}
\frac{{\rm d} E}{{\rm d} X} = \left(\frac{{\rm d} E}{{\rm d} X}\right)_{X_{\rm max}} \left(\frac{X-X_0}{X_{\rm
max}-X_0}\right)^{(X_{\rm max}-X_0)/\lambda} e^{\frac{X_{max}-X}{\lambda}} \,, \label{eq:gh}
\end{equation}
\noindent where $X_0$ and $\lambda$ are shape parameters and $X_{\rm max}$ is the shower maximum depth. 
\par
We have studied the impact of the FY selection for proton and iron showers with zenith angle $\theta$ of 30$^{\rm o}$ and 60$^{\rm o}$ and fixed primary energies $E_0$ of 10$^{19}$ and 10$^{20}$ eV. The corresponding calorimetric energies have been calculated assuming an invisible energy as parameterized by~\cite{song}. 
The values of $X_0$, $\lambda$ and $X_{\rm max}$ have been obtained by fitting the average longitudinal development of a sample of CORSIKA showers~\cite{perrone} to eq.~(\ref{eq:gh}). These longitudinal developments are fully compatible with those shown in \cite{SongLongitudinal}. The value of $\left({\rm d}E/{\rm d} X\right)_{X_{\rm max}}$ 
is obtained as a normalization constant to account for the total deposited energy. 
\par
\begin{figure}[htb]
\includegraphics[width=1.0\textwidth]{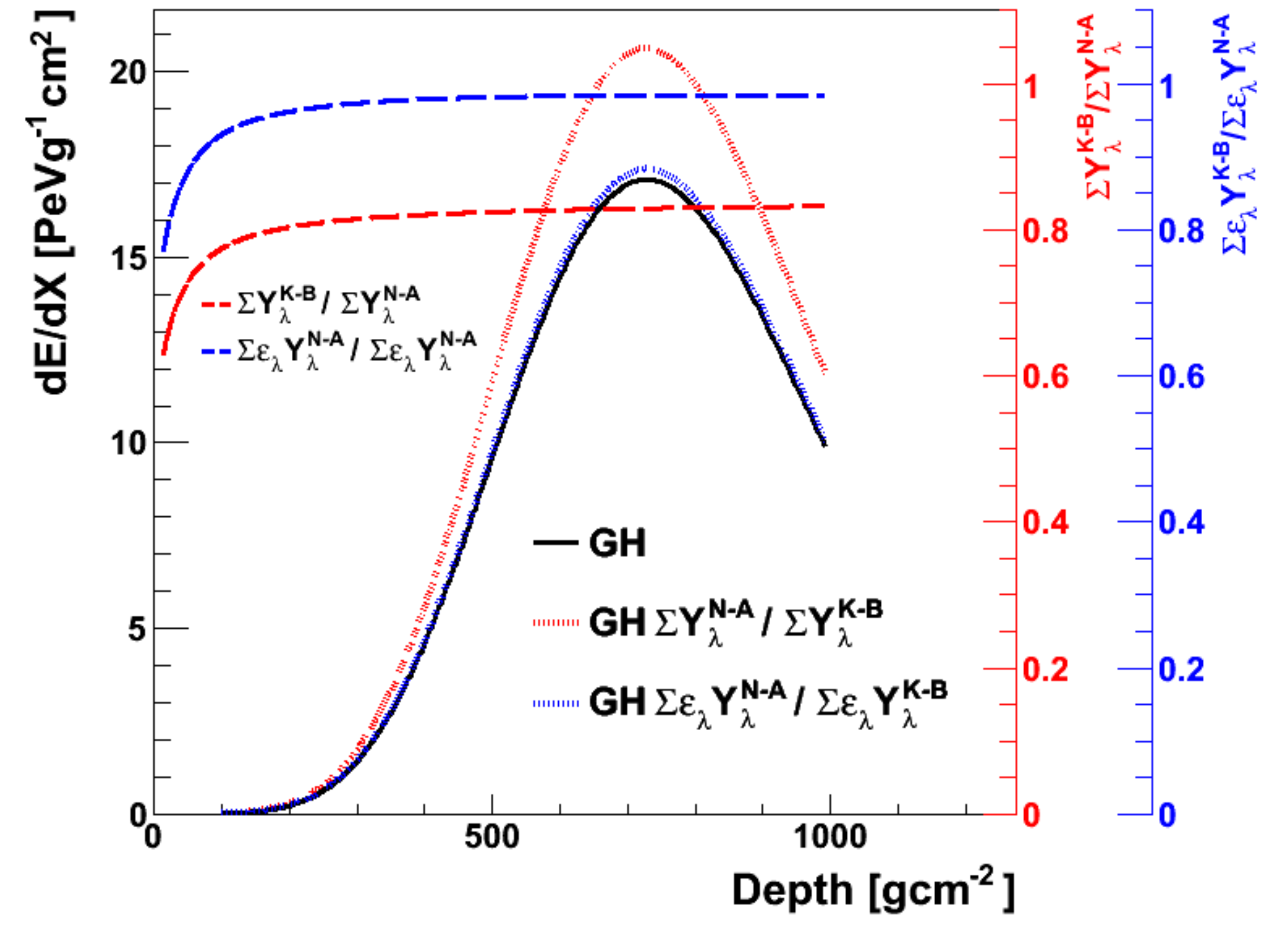}
\caption{\footnotesize{Comparison of FY N-A and K-B descriptions and their impact in the shower reconstruction for
a $E=10^{19}$ eV proton shower with 30$^o$ zenith angle (black profile).
The total FY of K-B and N-A (red axis on the right) are significantly different and therefore with a large impact on the profile of deposited energy (red profile). When including the optical efficiency of the telescopes the FY ratio (blue axis on the right) is very close to one and their impact on deposited energy in nearly negligible (blue profile).}}\label{fig:gh_1}
\end{figure}
\par
For a given energy, the shower reaches its maximum
development at an altitude which grows with $\theta$ and therefore, since the FY varies with altitude (through pressure,
temperature and humidity), the impact of the FY selection depends on the shower geometry. The effect of atmospheric transmission
increases with the distance between the telescope and the shower axis. Values of 10 and 30 km have been tried for this parameter. As already mentioned
the atmosphere has been modeled using monthly average profiles measured at the Auger site~\cite{keilhauer_atm}.
\par
As an example we show in figure~\ref{fig:gh_1} the effect in the profile of deposited  energy of a 10$^{19}$ eV proton shower
when changing from $Y^{\rm N-A}_\lambda(X)$ to $Y_\lambda ^{K-B}(X)$ fluorescence yield. The black solid line
is the GH profile for a shower with an $E_0$ value of $10^{19}$~eV. The red dashed line represents the ratio
$Y^{\rm K-B}_\lambda/Y_\lambda ^{\rm N-A}$ (red axis on the right) as a function of depth. Applying (\ref{eq:new_dep}), the
modified profile (i.e., the profile of deposited energy which would be obtained using the K-B fluorescence yield), has been
calculated (red dotted line). The integral of this profile results in a total deposited energy significantly larger (by about a 20\%). However when taking into account the effect of the optical efficiency, the ratio $\sum Y^{\rm
K-B}_\lambda(X)\cdot \varepsilon_\lambda /\sum Y^{\rm N-A}_\lambda(X)\cdot \varepsilon_\lambda$ is close to unity
except at very low atmospheric depth (dashed blue line measured in the blue axis on the right) and therefore the
reconstructed profile from expression (\ref{eq:new_dep_eff}) (dotted blue) is nearly unaffected giving a deviation in the
deposited energy around 2\% in full agreement with previous studies~\cite{unger}. The effect of the atmospheric transmission does not introduce significant differences between FY data sets in this case. This example has been calculated using the atmospheric profile of December. Similar results have been found for other seasons.

\begin{figure}[htb]
\includegraphics[width=0.49\textwidth]{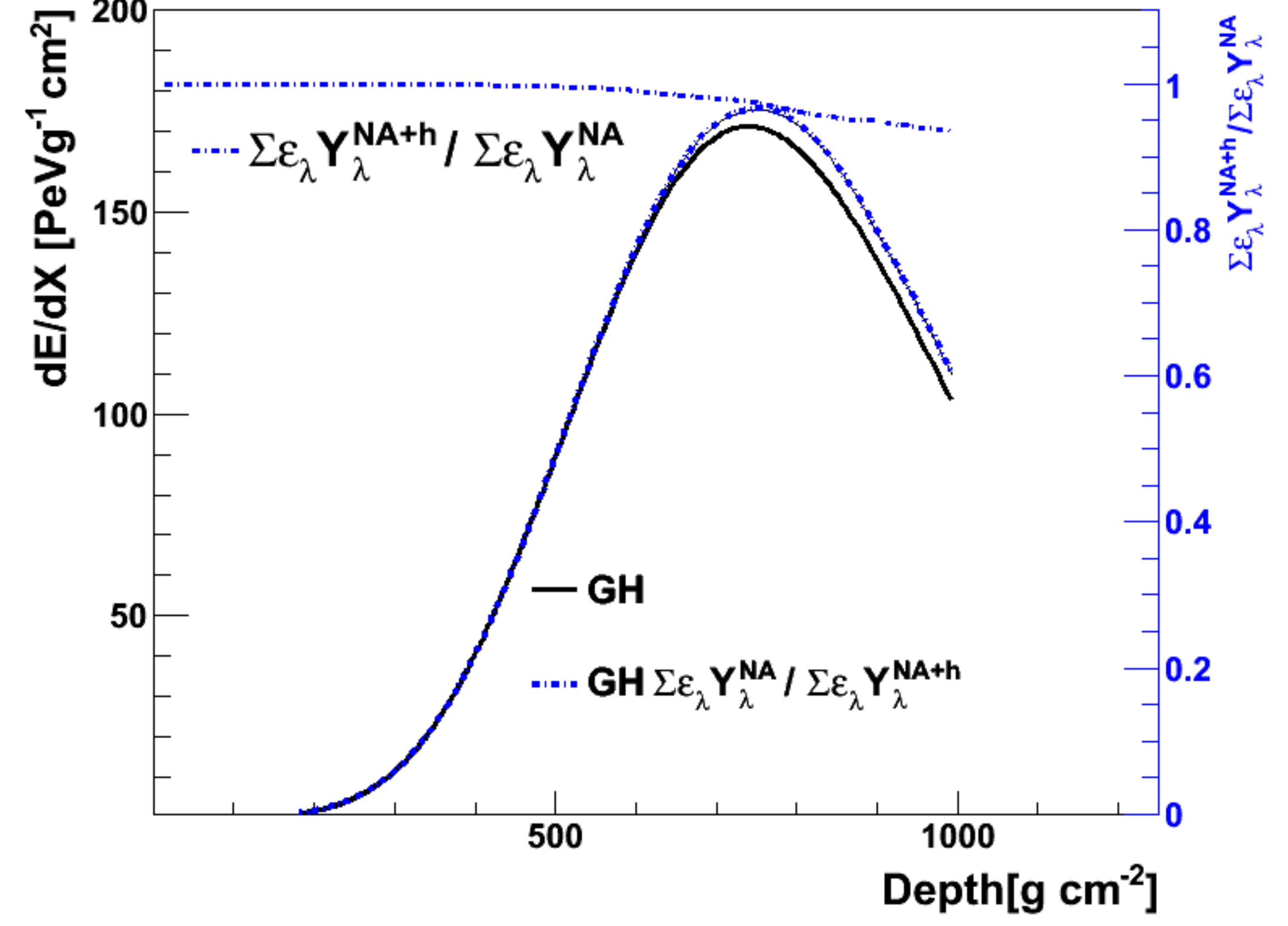}
\includegraphics[width=0.49\textwidth]{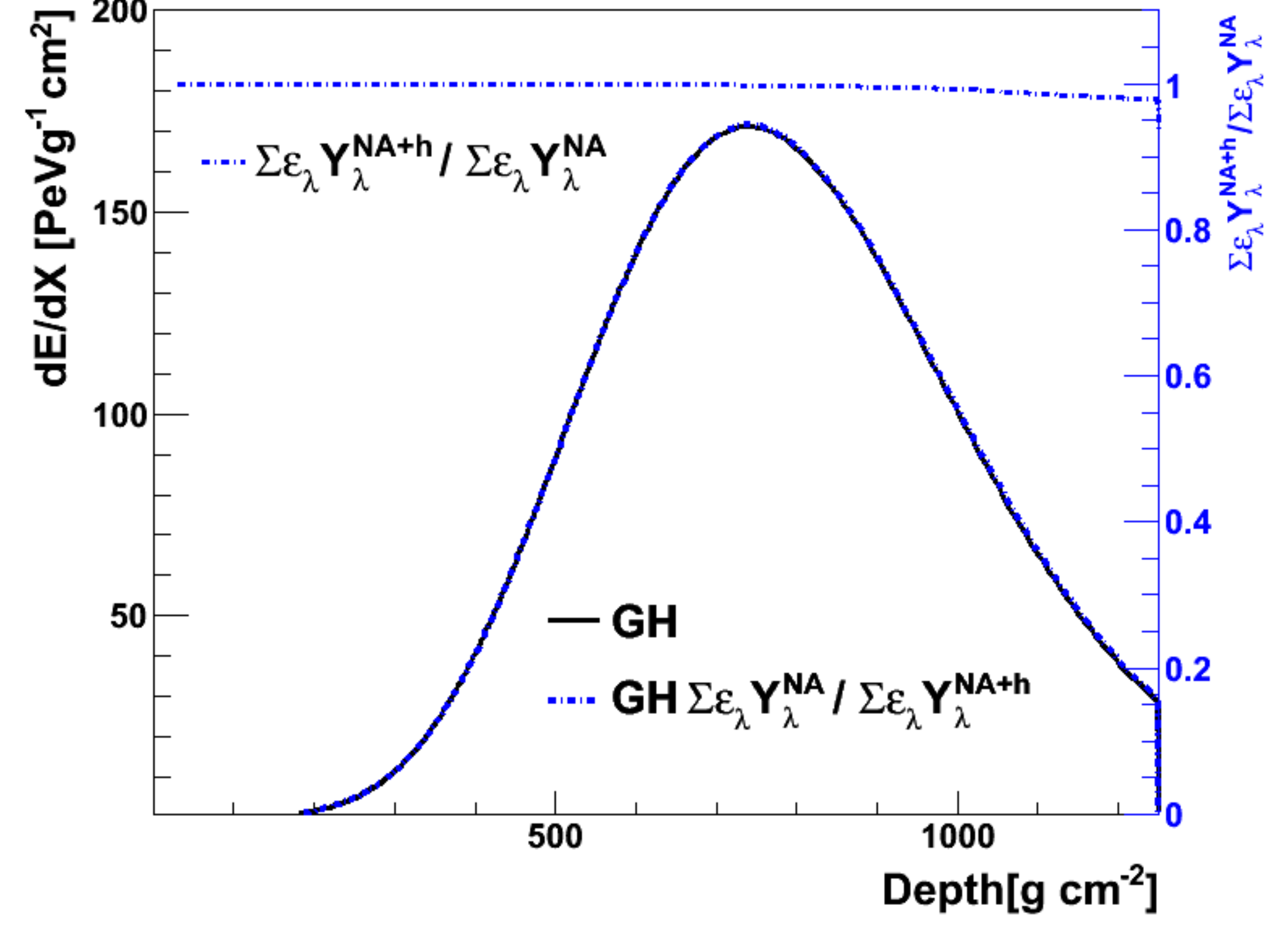}
\caption{\footnotesize{Effect of including the humidity contribution in the profile of energy deposited for 30$^{\rm o}$ (left) and
60$^{\rm o}$ (right) zenith angles for a Fe shower of 10$^{20}$ eV. Notice that the dependence of the FY ratio with slant depth varies with the incoming angle. See text for details.}}\label{fig:gh_hum}
\end{figure}
\begin{figure}[htb]
\includegraphics[width=0.49\textwidth]{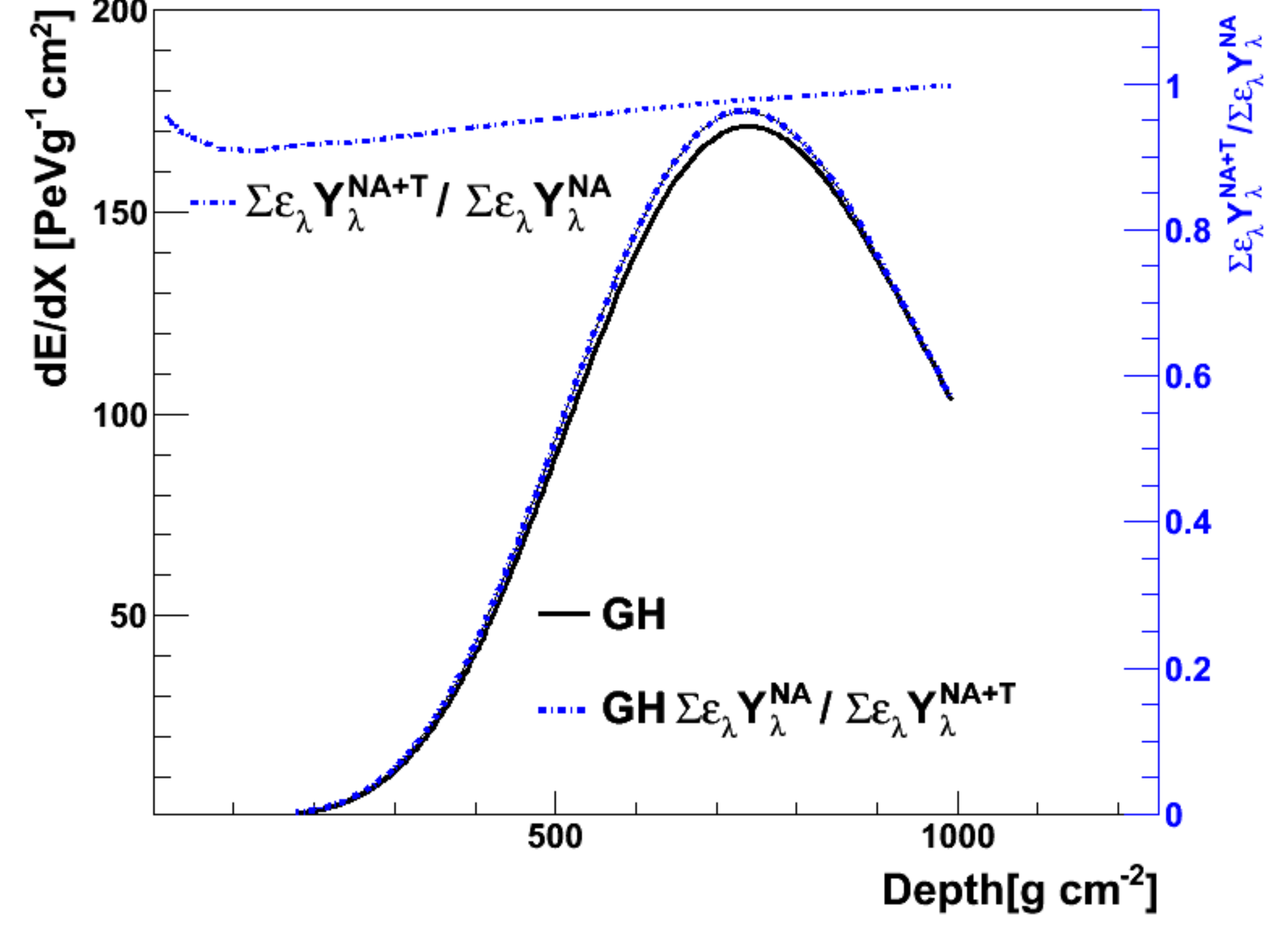}
\includegraphics[width=0.49\textwidth]{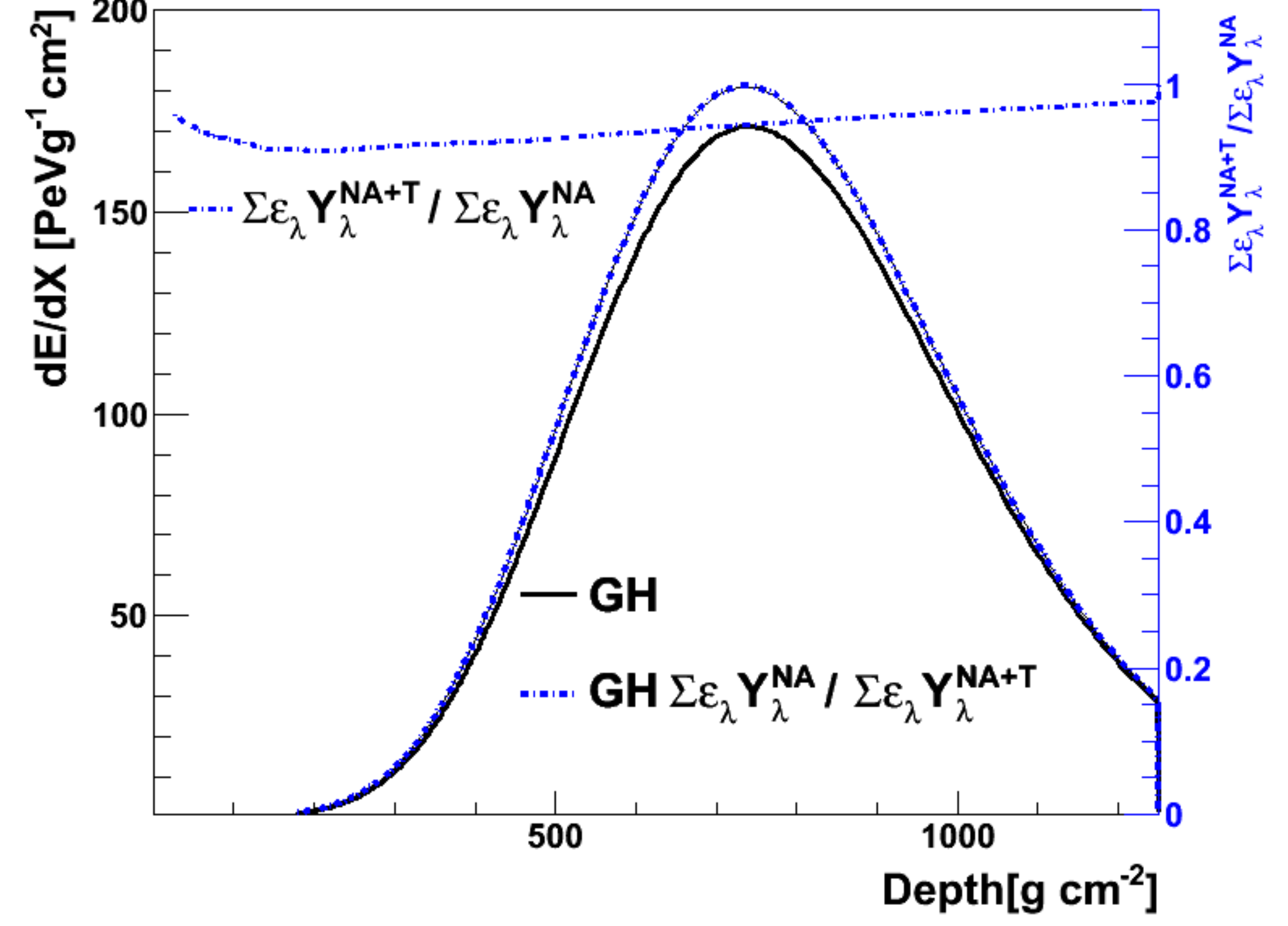}
\caption{\footnotesize{Effect of including the temperature dependence of the cross-section in the shower reconstruction for
30$^{\rm o}$ (left) and 60$^{\rm o}$ (right) zenith angles for the same case of figure~\ref{fig:gh_hum}.
See text for details.}}\label{fig:gh_temp}
\end{figure}
The effect of humidity and that of a $T$-dependent collisional cross section has also been studied with this method. Some
typical examples are shown in figures \ref{fig:gh_hum} and \ref{fig:gh_temp}. In figure~\ref{fig:gh_hum} the effect of water
vapor (December month) in the N-A data set for a Fe shower of $10^{20}$ eV and $\theta$ angles of 30$^{\rm o}$ and 60$^{\rm o}$
is studied. The development of the shower depends on the slant depth and therefore the smaller is $\theta$, the closer to the
ground the maximum shower development is reached. Since humidity is stronger at low altitudes, its effect is more important for
vertical showers. In this case the calorimetric energy has to be increased by about a 4\% at 30$^{\rm o}$ while at 60$^{\rm o}$ the
correction is negligible ($<$ 1\%). On the contrary the effect of neglecting the $T$ dependence of the collisional cross section
is stronger at low temperatures (i.e. at high altitudes) since the $T_0$ value at which the FY is measured are close to that at
ground. This feature can be observed in figure~\ref{fig:gh_temp} for the same kind of shower assuming the same atmospheric
profile. In this case the effect of temperature can be corrected by increasing
$E$ by about 2\% (6\%) at 30$^{\rm o}$ (60$^{\rm o}$).
\par
The corresponding effect of the correction on the shower maximum depth is also displayed in the above figures. As expected the $X_{\rm max}$ value increases (decreases) if the slope of the
FY ratio versus depth function at $X = X_{\rm max}$ is negative (positive).

The longitudinal profiles have been calculated until the showers reaches ground level that in the case of the Auger site corresponds to a vertical atmospheric depth of about 880 gcm$^{-2}$.

\section{Results}
\label{sec:results}
The method described above has been used to predict the impact of various FY assumptions on the shower reconstruction. 
The relative difference in reconstructed primary energy between several FY selections and the N-A one $\delta_E = (E_0-E_0^{\rm N-A})/E_0^{\rm N-A}$ has been calculated as well as the absolute difference of the shower maximum depth $\Delta X_{\rm max} = X_{\rm max} - X_{\rm max}^{\rm N-A}$. The effect of the optical efficiency of the telescope~\cite{auger_filter} (in particular that of the filter) and the atmospheric absorption to these deviations has also been evaluated. 
\par
As already mentioned the contribution of the invisible energy has been calculated following the parameterization of~\cite{song}. 
Since typical variations in deposited energy due to the FY selection are smaller than 10\% and the invisible energy contributes an amount ranging from 8\% and 12\%, the corresponding deviations in $E_0$ are smaller than the one of the calorimetric energy by as much as one percent unit.   
\par
In the first place the FY data sets of K-B and Nagano have been compared with the N-A one. The  results for total energy are displayed in tables~\ref{table_1} (10$^{20}$ eV) and \ref{table_2} (10$^{19}$ eV) while those of $X_{\rm max}$ are shown in tables~\ref{table_3} 
(10$^{20}$ eV) and \ref{table_4} (10$^{19}$ eV). As can be seen in tables~\ref{table_1} and \ref{table_2}, the Nagano FY would lead to a small decrease in energy (around or smaller than 2\%) without any additional relevant effect when the optical efficiency of the fluorescence telescope or the atmospheric transmission is included in the calculation. 
On the contrary, using the K-B FY would give rise to a significant energy increase of nearly 20\%. However, as anticipated in the previous
section, this deviation is reduced down to about 2\% when the optical efficiency of other components are included as pointed out by~\cite{unger}.
In particular the 
filter reduces this nearly 20\% energy increase down to about 7\%, with the optical efficiency of the remaining
components of the telescope responsible for the further decrease down to that final 2\%. The effect of the atmospheric
transmission is basically negligible
\begin{table}
\centering
\begin{tabular}{|c|c|c|c|c|c|c|c|c|}
\hline
\multicolumn{1}{|c|}{}  &               \multicolumn{4}{c|}{K-B/N-A}    &                 \multicolumn{4}{c|}{Nagano/N-A} \\\hline
\multicolumn{1}{|c|}{}  &  \multicolumn{2}{c|}{$30^{\rm o}$} &  \multicolumn{2}{c|}{$60^{\rm o}$} & \multicolumn{2}{c|}{$30^{\rm o}$} &  \multicolumn{2}{c|}{$60^{\rm o}$}  \\\hline
\multicolumn{1}{|c|}{}                                         &  p  &   Fe   &   p   &   Fe  &   p  &   Fe &  p   &   Fe\\\hline
$Y_\lambda(X)$                                                 & 19  &   19   &  19   &  20   &  -1  & -1   & -1   & -1    \\\hline
$Y_\lambda(X)\cdot F_\lambda$                                  & 7   &   7    &   7   &   8   &   0  &  0   &  0   &  0    \\\hline
$Y_\lambda(X)\cdot \varepsilon_\lambda$                           & 2   &   2    &   2   &   3   &  -2  & -2   & -1   & -1     \\\hline
$Y_\lambda(X)\cdot \varepsilon_\lambda \cdot T_\lambda(X)$ (10km) & 1   &   1    &   1   &   2   &  -2  & -2   & -1   & -1     \\\hline
$Y_\lambda(X)\cdot \varepsilon_\lambda \cdot T_\lambda(X)$ (30km) & 0   &   0    &   0   &   0   &  -2  & -2   & -2   & -2    \\\hline
\end{tabular}
\caption{\footnotesize{Percentage differences (100$\delta_E$) between the primary energy reconstructed using either K-B or Nagano instead of the N-A FY data set, for typical GH profiles of p and iron $10^{20}$ eV showers. These results have been obtained using the atmospheric
profile of December. The effect of the filter $F_{\lambda}$, the total optical efficiency $\varepsilon_{\lambda}$
and the atmospheric transmission $T_{\lambda}$ assuming shower-telescope distances of 10 and 30 km is shown. Uncertainties in 100$\delta_E$ are
estimated in about $\pm$0.5 and thus null results represent energy deviations below 0.5\%}}
\label{table_1}
\end{table}
\begin{table}
\centering
\begin{tabular}{|c|c|c|c|c|c|c|c|c|}
\hline
\multicolumn{1}{|c|}{}  &               \multicolumn{4}{c|}{K-B/N-A}    &                 \multicolumn{4}{c|}{Nagano/N-A} \\\hline
\multicolumn{1}{|c|}{}  &  \multicolumn{2}{c|}{$30^{\rm o}$} &  \multicolumn{2}{c|}{$60^{\rm o}$} & \multicolumn{2}{c|}{$30^{\rm o}$} &  \multicolumn{2}{c|}{$60^{\rm o}$}  \\\hline
\multicolumn{1}{|c|}{}                                         &  p   &   Fe   &   p   &   Fe  &   p  &   Fe  &  p    &   Fe\\\hline
$Y_\lambda(X)$                                                 & 19   &  19    &  20   &  20   &  -1  & -1    & -1    & -1    \\\hline
$Y_\lambda(X)\cdot F_\lambda$                                  &  7   &   7    &   8   &  8    &   0  &  0    &  0    &  0    \\\hline
$Y_\lambda(X)\cdot \varepsilon_\lambda$                           &  2   &   2    &   3   &  3    &  -2  & -2    & -1    & -1     \\\hline
$Y_\lambda(X)\cdot \varepsilon_\lambda \cdot T_\lambda(X)$ (10km) &  1   &   1    &   2   &  2    &  -2  & -2    & -1    & -1     \\\hline
$Y_\lambda(X)\cdot \varepsilon_\lambda \cdot T_\lambda(X)$ (30km) &  0   &   0    &   0   &  0    &  -2  & -2    & -2    & -2     \\\hline
\end{tabular}
\caption{\footnotesize{Same as table \ref{table_1} for $10^{19}$ eV.}}
\label{table_2}
\end{table}
\begin{table}
\centering
\begin{tabular}{|c|c|c|c|c|c|c|c|c|}
\hline
\multicolumn{1}{|c|}{}  &               \multicolumn{4}{c|}{K-B/N-A}    &                 \multicolumn{4}{c|}{Nagano/N-A} \\\hline
\multicolumn{1}{|c|}{}  &  \multicolumn{2}{c|}{$30^{\rm o}$} &  \multicolumn{2}{c|}{$60^{\rm o}$} & \multicolumn{2}{c|}{$30^{\rm o}$} &  \multicolumn{2}{c|}{$60^{\rm o}$}  \\\hline
\multicolumn{1}{|c|}{}                                         &   p  &   Fe &    p  &   Fe  &   p  &    Fe &  p    &   Fe    \\\hline
$Y_\lambda(X)$                                                 &  -1  &  -1  &   -2  &  -2   &  -1  &  -1   &  -1   &  -1     \\\hline
$Y_\lambda(X)\cdot F_\lambda$                                  &  -1  &  -1  &   -2  &  -2   &  -1  &  -1   &  -1   &  -1     \\\hline
$Y_\lambda(X)\cdot \varepsilon_\lambda$                           &  -1  &  -1  &   -2  &  -1   &   0  &   0   &  -1   &  -1     \\\hline
$Y_\lambda(X)\cdot \varepsilon_\lambda \cdot T_\lambda(X)$ (10km) &   0  &   0  &   -1  &  -1   &   0  &   0   &  -1   &  -1     \\\hline
$Y_\lambda(X)\cdot \varepsilon_\lambda \cdot T_\lambda(X)$ (30km) &   1  &   0  &   -2  &  -2   &   0  &   0   &  -1   &  -1     \\\hline
\end{tabular}
\caption{\footnotesize{$\Delta X_{\rm max}$ differences (gcm$^{-2}$) between the reconstructed values using K-B or Nagano
instead the N-A data set, for typical GH profiles of p and iron $10^{20}$ eV showers. Uncertainties in
$\Delta X_{\rm max}$ are estimated in about $\pm$0.5 gcm$^{-2}$. Null results represent absolute deviations below 0.5gcm$^{-2}$.
More details on first column are given in table \ref{table_1}.}} \label{table_3}
\end{table}
\begin{table}
\centering
\begin{tabular}{|c|c|c|c|c|c|c|c|c|}
\hline
\multicolumn{1}{|c|}{}  &               \multicolumn{4}{c|}{K-B/N-A}    &                 \multicolumn{4}{c|}{Nagano/N-A} \\\hline
\multicolumn{1}{|c|}{}  &  \multicolumn{2}{c|}{$30^{\rm o}$} &  \multicolumn{2}{c|}{$60^{\rm o}$} & \multicolumn{2}{c|}{$30^{\rm o}$} &  \multicolumn{2}{c|}{$60^{\rm o}$}  \\\hline
\multicolumn{1}{|c|}{}                                         &   p  &   Fe   &   p   &   Fe  &   p  &    Fe &  p    &   Fe    \\\hline
$Y_\lambda(X)$                                                 &  -1  &    -1  &   -2  &  -2   &  -1  &  -1   &  -1   &   1     \\\hline
$Y_\lambda(X)\cdot F_\lambda$                                  &  -1  &    -1  &   -2  &  -2   &  -1  &  -1   &  -1   &  -1     \\\hline
$Y_\lambda(X)\cdot \varepsilon_\lambda$                           &  -1  &    -1  &   -2  &  -2   &   0  &  -1   &  -1   &  -1     \\\hline
$Y_\lambda(X)\cdot \varepsilon_\lambda \cdot T_\lambda(X)$ (10km) &   0  &    -1  &   -1  &  -1   &   0  &  -1   &  -1   &  -1     \\\hline
$Y_\lambda(X)\cdot \varepsilon_\lambda \cdot T_\lambda(X)$ (30km) &   0  &     0  &   -2  &  -2   &   0  &   0   &  -1   &  -1     \\\hline
\end{tabular}
\caption{\footnotesize{Same as table \ref{table_3} for $10^{19}$ eV.}}
\label{table_4}
\end{table}
\par
In regard to $X_{\rm max}$ reconstruction, using the FY data sets of Nagano or K-B would decrease the shower
maximum depth by about 1 gcm$^{-2}$ (see tables~\ref{table_3} and \ref{table_4}). The effect of the atmospheric transmission is nearly negligible. These results both of reconstructed energy and maximum depth are basically independent on primary energy and shower angle.
As expected, they are also nearly independent of season.
\par
The effect of the humidity and the $T$-dependent collisional cross section on the N-A data set has also been studied. 
As already mentioned these effects vary with geometry and season due to the different atmospheric conditions undergone by the shower track, as well as with the
nature of the primary (mass and energy) due to the different longitudinal development.
\par
\begin{figure}
\includegraphics[height=0.6\textwidth]{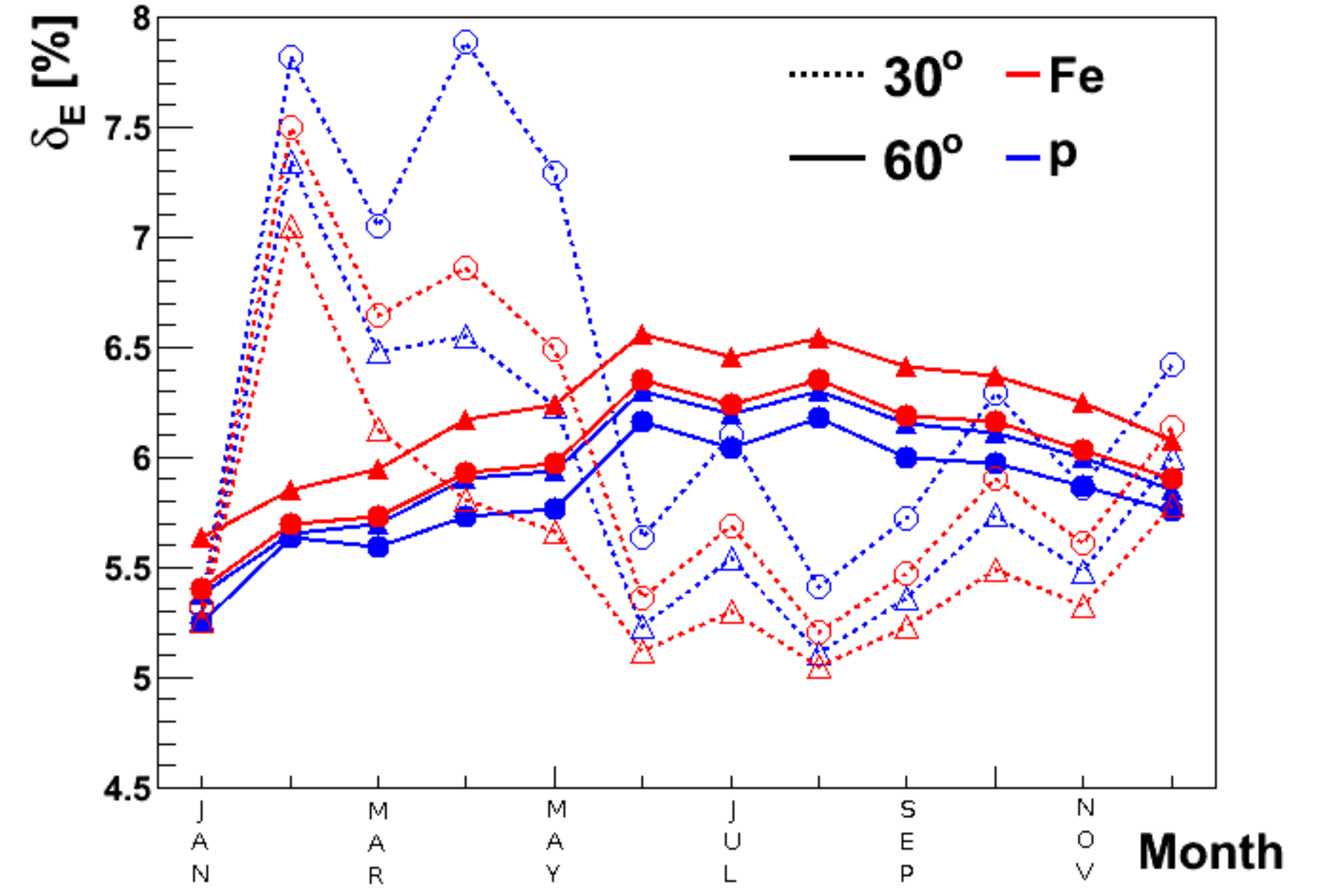}
\caption{\footnotesize{Percentage increase in reconstructed $E_0$ value when including humidity and a collisional cross section dependent on temperature to the N-A data set, versus month. Results are shown for both protons and Fe nuclei of $10^{19}$ eV (triangles) and $10^{20}$ eV (circles) energies.}}
\label{fig:delta_E}
\end{figure}

When the water vapour quenching is taken into account, large deviations in the calorimetric energy (up to 7\%) are found for humid months (e.g. February) and shower maximum close to ground (i.e., either high energy, light primary, vertical incidence or a combination of them). This deviation lowers down to about 1\% at 60$^{\rm o}$. The effect of a collisional cross-section dependent on temperature ranges from 6.5\% for inclined showers to about 2\% for vertical ones. The total effect in the primary energy is displayed for all months in figure~\ref{fig:delta_E}. Excluding the February - May period, it ranges typically between 5.5 and 6.5\%. Notice the significant variation of $\delta_E$ with season for vertical showers, which is due to the humidity effect.

\begin{figure}
\includegraphics[height=0.6\textwidth]{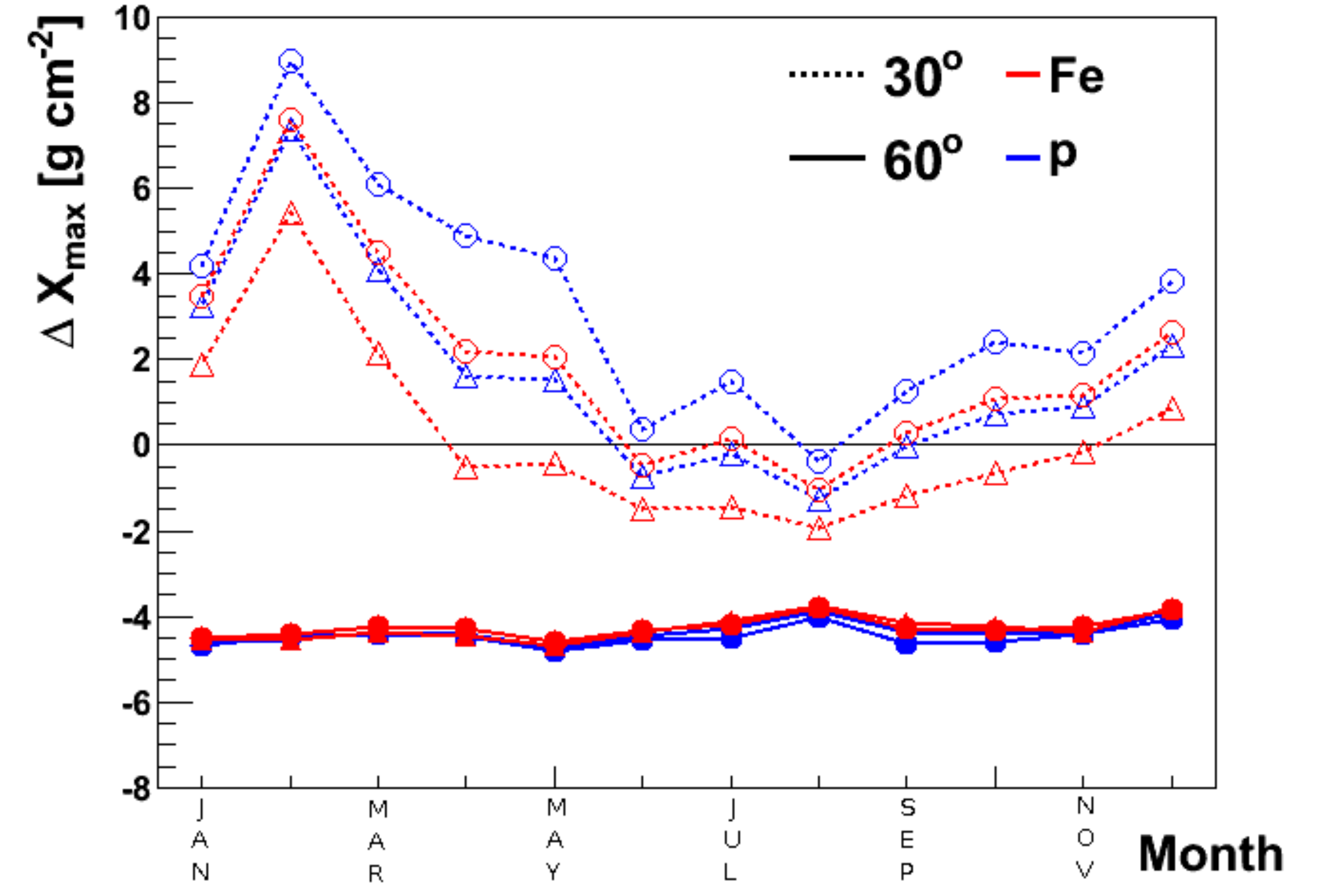}
\caption{\footnotesize{Increase of shower maximum depth due to humidity and a collisional cross section dependent on temperature to the N-A data set, versus month. See figure \ref{fig:delta_E} for more details.}}
\label{fig:Delta_X}
\end{figure}
\par
The correction in shower maximum depth due to the temperature effect is $\Delta X_{\rm max} \approx$-5 gcm$^{-2}$, nearly independent on the shower and atmospheric features while the correction for the humidity effect ranges from about 8 gcm$^{-2}$ for vertical showers in austral summer down to about 1 gcm$^{-2}$ at 60$^{\rm o}$. The total $X_{\rm max}$ deviations including both effects are displayed for all months in figure~\ref{fig:Delta_X}. As can be seen in this figure, the corrections for inclined showers are negative and nearly season independent. This behaviour was expected since the slope of the function FY ratio versus depth is positive for the $T$ effect (see figure~\ref{fig:fy_hT}) and, as already mentioned, the temperature profile is nearly constant along the year. On the contrary the correction for vertical showers is strongly dependent on the month again due to the humidity effect. 
\par
The experimental analysis of the fluorescence quenching due to water vapor as well as that of the temperature dependence of the quenching cross section is rather difficult. As a consequence these measurements are very scarce and subject to large uncertainties. The procedure presented here can be used for the evaluation of the effects of the errors in the $P'_{\rm w}$ and $\alpha$ parameters on the total uncertainty of the reconstructed values of the shower energy and $X_{\rm{max}}$. For this purpose we have recalculated $\delta E$ and $\Delta X_{\rm max}$ after increasing (decreasing) $\alpha_{\lambda}$ by a factor 1.5 (0.5) and increasing (decreasing) $P'_{\rm w}$ by a factor 1.2 (0.8) for all molecular bands. In general the effect of such large eventual deviations in these parameters is very small. The largest effect is found for high humidity months and/or showers more sensitive to $T$ and humidity effects.
\par
For instance for 30$^{\rm o}$ protons of 10$^{20}$ eV the 8\% increase in energy for the February month 
would move within the 6\% to 10\% interval while the corresponding shift in maximum depth of $\Delta X_{\rm max}$ = 9 gcm$^{-2}$ 
remains unchanged. For the same kind of shower (p, 30$^{\rm o}$, 10$^{20}$ eV)  the 5\% energy increase in the August month would be within the 3 to 6\% interval and 
the negligible $X_{\rm max}$ effect reported in fig.~\ref{fig:Delta_X}∫® would be in between -2 and -1 gcm$^{-2}$. Finally for Fe showers of 60$^{\rm o}$ and 10$^{20}$ eV the
6\% energy increase in August would be in between 3 and 9\% and the corresponding $\Delta X_{\rm max}$ = -4 gcm$^{-2}$ value would be within the -5 to -1 gcm$^{-2}$
interval. Note that the above examples represent the cases with the largest dependence on possible uncertainties in the $P'_{\rm w}$ and $\alpha$ parameters.
\section{Conclusions}
\label{sec:conclusions} A simple analytical method has been proposed to quantify the influence of the assumed FY on the shower reconstruction, in particular on primary energy and shower maximum depth. The longitudinal development of the shower is described by a Gaisser-Hillas profile which is slightly modified by a variation in the FY assumption. Several data sets of FY including absolute values, wavelength spectra, as well as pressure, temperature and humidity dependencies have been used for this study. 
\par
Using this simple procedure we have confirmed that the Nagano-AIRFLY, Nagano and Kakimoto-Bunner data sets lead to close values (within around a 2\%) of the reconstructed energy, as long as the optical efficiency of the Auger telescopes is taken into account. The corresponding effects on $X_{\rm max}$ are typically smaller than 1 gcm$^{-2}$. 
\par
The dependence of the FY with atmospheric properties ($P$, $T$ and humidity) and its effect on reconstructed parameters of the shower have also been analyzed. These effects, when combined, introduce typical deviations at the level of 6\% in the reconstructed primary energy. However significantly larger deviations can be found for vertical showers (30$^{\rm o}$) in humid months. The deviations induced in $X_{\rm max}$ by temperature and humidity are opposite and they nearly cancel when combined. Deviations of about -4 gcm$^{-2}$ are found for inclined showers (60$^{\rm o}$) while for vertical ones the effect is strongly dependent on humidity with large positive deviations in some months. 
\par
These deviations in primary energy and $X_{max}$ values are not negligible when compared with typical uncertainties (both systematic and statistical) in the reconstruction of the shower parameters. Notice, however, that they can be easily corrected as far as the temperature and humidity effects are included in the reconstruction algorithms. On the other hand, we have found that even relatively large uncertainties in the quenching parameters associated to humidity or the T-dependence of the collisional cross section have not a significant impact on the reconstructed parameters. In fact the derived systematic uncertainties are below 2\% for primary energy and below 2gcm$^{-2}$ for $X_{max}$, and therefore they represent a small contribution to the total uncertainty in the reconstruction of the shower parameters.
\par
The results shown here have been obtained from the whole longitudinal development of the fluorescence light. In a real case the reconstruction 
of the longitudinal development is restricted to an interval given by the field of view of the telescope. In addition, the contribution of Cherenkov light is a very important ingredient  in the reconstruction of real data \cite{unger_2} not used in our simple procedure. These simplifications could give rise to small 
discrepancies between the predictions of our simple algorithm and those from the analysis of real data. Nevertheless our results are basically in agreement with those of \cite{ICRC_09_Bianca} obtained from a detailed reconstruction of a sample of MC events.
\par

\pagebreak
\section*{Acknowledgements}
This work has been supported by the Spanish Ministerio de Ciencia e Innovacion (FPA2009-07772 and CONSOLIDER CPAN CSD2007-42) and ``Comunidad de Madrid" (Ref.: 910600). M.~Monasor acknowledges the ``Consejer\'ia de Educaci\'on y Ciencia de Castilla-La Mancha'' and the ``Fondo Social Europeo'' for a postdoctoral fellowship. Very fruitful discussion with our colleagues of the Auger collaboration are acknowledged, in particular with B. Keilhauer, J. Matthews, V.~H.~Ponce and M. Unger.


\begin{thebibliography}{99}
\bibitem{ICRC_09_FY} M.~Monasor \etal, ``The impact of the fluorescence yield on the reconstructed shower parameters of ultra-high energy cosmic rays'', Proc. 31st ICRC, Lodz, Poland (2009).

\bibitem{ICRC_09_Bianca} B.~Keilhauer and M.~Unger, ``Fluorescence emission induced by extensive air showers in dependence on atmospheric conditions",  Proc. 31st ICRC, Lodz, Poland (2009).

\bibitem{5th_FW_SP} F.~Arqueros \etal, Proc. 5th Fluorescence Workshop,
El Escorial, Madrid, Nucl. Instr. and Meth. A {\bf 597} (2008) 1.

\bibitem{rosado} J.~Rosado \etal, Astropart. Phys. {\bf 34} (2010) 164.

\bibitem{Tilo} T.~Waldenmaier \etal, Astropart. Phys. {\bf 29} (2008) 205.

\bibitem{temp_cross_sec} M.~Ave \etal [AIRFLY Collaboration], Proc. 5th Fluorescence Workshop,
El Escorial, Madrid, Nucl.\ Instrum.\ Meth.\ A {\bf 597} (2008) 50.

\bibitem{cross_section}D.~L.~Holtermann {\it et al.}, J. Chem. Phys. {\bf 77} (1982) 5327.

\bibitem{NIM_pioneering} F.~Arqueros \etal, Proc. 5th Fluorescence Workshop,
El Escorial, Madrid, Nucl. Instr. and Meth. A {\bf 597} (2008) 23.

\bibitem{dawson} B.~Dawson, Private Communication.

\bibitem{kakimoto} F.~Kakimoto \etal, Nucl. Instr. and Meth. A {\bf 372} (1996) 527.

\bibitem{bunner_thesis} A.~N.~Bunner, ``Cosmic Ray Detection by Atmospheric Fluorescence'', Ph.D. Thesis, Cornell University, Ithaca, N.Y. (1967).

\bibitem{nagano} M.~Nagano, \etal, Astropart. Phys. {\bf 22} (2004) 235.

\bibitem{moriond} F.~Arqueros \etal, 43rd Rencontres de Moriond: Electroweak Interactions and Unified Theories, La Thuile, Italy (2008) (arXiv:0807.4824); F.~Arqueros \etal, New J. Phys. {\bf 11} (2009) 065011. 

\bibitem{njp_nagano} M.~Nagano, New J. Phys. {\bf 11} (2009) 065012.

\bibitem{AIRFLY_2007} M.~Ave \etal [AIRFLY Collaboration], Astropart. Phys. {\bf 28} (2007) 41.

\bibitem{auger_spectrum} J.~Abraham \etal [Pierre Auger Collaboration], Physics Letters B {\bf 685} (2010) 239  

\bibitem{auger_filter} J.~Abraham \etal [Pierre Auger Collaboration], Nucl.\ Instrum.\ Meth.\  A {\bf 620} (2010) 227.

\bibitem{keilhauer_atm} J. Abraham \etal [Pierre Auger Collaboration], Astropart. Phys. {\bf 33} (2010) 108. 

\bibitem{bianca_NIM_A} B.~Keilhauer \etal, Proc. 5th Fluorescence Workshop, El Escorial, Madrid, Nucl.\ Instrum.\ Meth.\ A {\bf 597} (2008) 99.

\bibitem{gaisser-hillas}T.~Gaisser and G.~Hillas, Proc. 15th International Cosmic Ray Conference, Plovdiv {\bf 8} (1977) 353.

\bibitem{song} C.~Song \etal, Astropart. Phys. {\bf 14} (2000) 7.

\bibitem{perrone} L.~Perrone, S. Petrera and F. Salamida. Private Communication.

\bibitem{SongLongitudinal} C.~Song. Astropart. Phys. {\bf 22} (2004) 151.

\bibitem{unger} F.~Sch\"ussler and M.~Unger, Private Communication.

\bibitem{unger_2} M.~Unger \etal, Nucl. Instr. and Meth. A{\bf 588} (2008) 433.

\end{thebibliography}
\end{document}